\documentclass[aip,jcp,reprint,superscriptaddress,floatfix, onecolumn]{revtex4-1}
\usepackage{graphicx}
\usepackage{amssymb}
\usepackage[separate-uncertainty=true,exponent-product=\!\cdot\!]{siunitx}
\usepackage{physics}
\graphicspath{{figures/}}
\usepackage[colorlinks=true,allcolors=blue]{hyperref}
\usepackage{physics}
\newcommand{\funcd}{\boldsymbol{\delta}}

\DeclareMathOperator{\sh}{sh}
\DeclareMathOperator{\ch}{ch}

\begin{document}

\title{Power efficiency of Hall-like devices: comparison between reciprocal and anti-reciprocal Onsager relations}
 \author{Jean-Eric Wegrowe} \email{jean-eric.wegrowe@polytechnique.edu}
\affiliation{LSI, \'Ecole Polytechnique, CEA/DRF/IRAMIS, CNRS, Institut Polytechnique de Paris, 91120 Palaiseau, France}
\author{Luqian Zhou}
\affiliation{LSI, \'Ecole Polytechnique, CEA/DRF/IRAMIS, CNRS, Institut Polytechnique de Paris, 91120 Palaiseau, France}
\author{Sariah Al Saati}
\affiliation{Centre de Physique Théorique, 91120 Palaiseau, France}
\date{\today}

\date{\today}

\begin{abstract}
 Two well-known Hall-like effects are occurring in ferromagnets: the Anomalous Hall effect and the Planar Hall effect.  The former is analogous to the classical Hall effect and is defined by the Onsager reciprocity relation of the second kind (antisymmetric conductivity matrix), while the latter is defined by the Onsager reciprocity relation of the first kind (symmetric conductivity matrix). The difference is fundamental, as it is based on time-invariance symmetry breaking at the microscopic scale. We study the Hall current generated in both cases, together with the power that can be extracted from the edges of Hall device.  The expressions of the distribution of the electric currents, the distribution of electric carriers, and the power efficiencies (i.e. the power that can be injected into a load circuit) are derived at stationary regime from a variational method based on the second law of thermodynamics. It is shown that the distribution of the transverse Hall-current is identical in both cases but the longitudinal current and the power dissipated differ at the second order in the Hall angle.
\end{abstract}

\maketitle

\section{Introduction}
 
Two hundred years ago exactly, in 1824, Sadi Carnot published a seminal work that contained the first formulation of the second law of thermodynamics \cite{Carnot}. Independently,  fifty five years letter, in 1879, Edwin Hall published a report about the effect that bears his name \cite{Hall}. And fifty two years later, in 1931, Lars Onsager published two papers  \cite{Onsager1,Onsager2} on the reciprocity relations that describe the cross-effects occurring in transport phenomena. This  breakthrough was crowned by a Nobel prize in 1968 \cite{Nobel_Lecture}. More recently, and related to the three discoveries recalled above, a series of fascinating studies have been devoted to the anomalous Hall effect (AHE), and its link with topological materials. This series started with a paper of Karplus and Luttinger in 1954 \cite{Luttinger}, ending fifty years latter \cite{Kondo,Nozieres, Berger, Bruno} with a paper of Haldane \cite{Haldane} and a Noble prize in 2016. This subject is still very productive today\cite{AHE}, especially within the context of new topological materials exhibiting unconventional AHE \cite{Altermag,Mazin}. \\

In a first paper, Onsager derived the reciprocity relation of the first kind (called reciprocal) from the microscopic time-reversal symmetry \cite{Onsager1}. The second paper deals with the reciprocity relation of the second kind (called antireciprocal) derived from the partial breaking of this time-reversal symmetry, typically generated by an axial vector\cite{Onsager2}. The first reciprocity relation leads to a symmetric matrix of the transport coefficients, while the second reciprocity relation leads to an antisymmetric transport matrix. 
Because of his simplicity, seniority, and ubiquity \cite{Dyakonov,Onoda}, Hall-like effects can be considered as an archetype of antireciprocal Onsager relations, applied to electric transport. The Hall-like effects are characterized by an antisymmetric conductivity matrix, for which the off-diagonal coefficients represent the coupling between the two directions $x$ and $y$ of the plane of the Hall device. In a ferromagnet (or unconventional magnets), two well-known Hall-like effects are observed: the Anomalous Hall effect (AHE) and the Planar Hall effect (PHE) \cite{PHE_Review,Schuhl}. In contrast to AHE, the PHE is due to the anisotropic magnetoresistance, and is defined by the reciprocal Onsager relation (symmetric conductivity matrix). Like for AHE, the lateral Hall-current produced by PHE can be used in order to switch an adjacent ferromagnetic layer\cite{Krivorotov} (an effect called spin-orbit torque, or SOT \cite{SOT}). 
In the present report, we study the electric power that can be injected into a load circuit through the Hall-current. We used the PHE as an archetype of the reciprocal Onsager symmetry relation, while the AHE is used as an archetype of the reciprocal relation.   
\\ 

 In the context of the theoretical studies about dissipation in the AHE, the discussion focused on the microscopic mechanisms responsible for the effect \cite{AHE}. Yet, at the macroscopic scale, dissipative transport phenomena do not follow the microscopic rules, {\it since there are first governed by the second law of thermodynamics}, for which the global constraints (the position of the power generators, the configuration of the contacts, the topology of the device, etc) play a major role \cite{Benda}.

The role of the second law of thermodynamics in the Hall effect has been overlooked \cite{Popovic,Putley} because, as pointed-out in reference Ref. \cite{Moelter}, the ``voltage differences between suitable pairs of points in the current flow suffice to characterize the transport processes''. However, such a characterization is incomplete because the Hall bar is not analogous to a capacitor. Indeed, a specificity of the Hall effect is that the accumulation of the electric carriers at the edges are produced by the system itself, in order to counter the action of the magnetic field on the electric carriers (Lenz law). The values of both the density and the current at the edges cannot be taken as independent local boundary conditions. An immediate consequence is that both the density of electric carriers and the current distribution inside a perfect Hall bar are not homogeneous.  More puzzling: Hall-like systems seem not to be reducible to a classical lumped-element circuit obeying the Kirchhoff's laws \cite{Gyrator}. Indeed, if it were the case, we would have the same lumped-element circuit for both PHE and AHE (both effects are described by the same set of parameters: resistances $R$ and $R_{\ell}$, Hall angle $\Theta$, and the  current $J_x^0$ injected by the generator, as sketched in Fig.1), which is in contradiction with the fact that the stationary states are different in both systems (as shown below). \\

In  previous works, a variational method based on the second law of thermodynamics \cite{Benda,JAP1,JAP2,JAP3,SHE}  has been developed in order to calculate the power dissipated by the Hall current in a perfect Hall bar in contact with a lead circuit \cite{JAP3}. Due to the small value of the Hall angle (below 1\%), the properties of the dissipated power are found to be surprisingly close to that expected for of a usual lumped-element circuit. The validity of the results have however been confirmed by an experimental study performed on the AHE in GdCo ferrimagnets \cite{ArXiv_2024}. \\

These results raised the question about the difference of stationary states generated between {\it reciprocal vs. antireciprocal} Onsager relations. The goal of the present work is to answer this question for AHE vs. PHE devices, and to present a synoptic calculation based on the stationary variational method, all other parameters being equivalent. It is shown that the transverse Hall-current  is identical in both cases, while the difference for the power is of the second order with respect to the Hall angle. These results confirm the observation that  the Joule dissipation is surprisingly similar - but not equal - for both AHE and PHE. In particular, both effects find the maximum of the efficiency for the resistance matching condition. The synoptic presentation proposed below allows the differences and similarities to be better understood. 

\section{Model}
 \begin{figure}[ht!]
          \centering
          \includegraphics[width=0.35\textwidth]{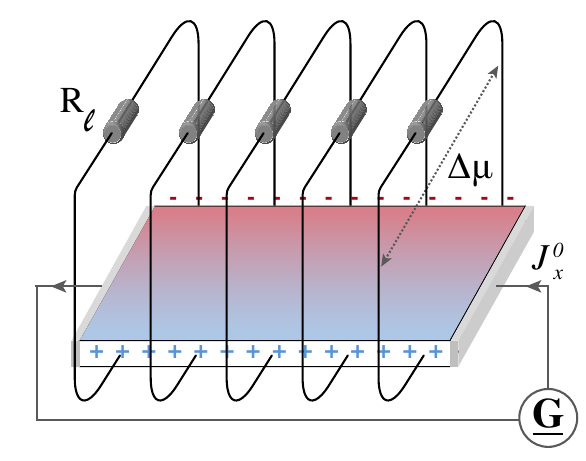}   \qquad
                 \includegraphics[width=0.35\textwidth]{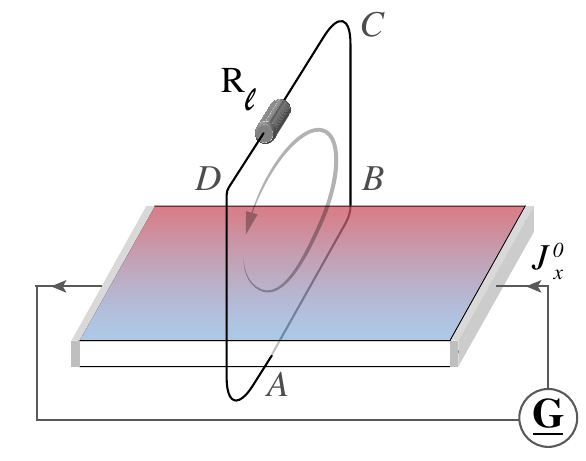}
         \caption{Left: sketch of the Hall bar contacted to a load circuit (resistance $R_\ell$) that preserves the translation invariance along $x$. The sign of the electric-charge accumulation is indicated by red (-) and blue (+) colors (see the calculated profile in Fig.3). Right: Integration loop ABCDA defined on one of the $yz$ vertical planes. A magnetic vector $\vec m$ is present but it is not represented in the picture (see definition of the Hall angles in the text).}
         \label{fig:opencicuit}
 \end{figure}

The system is a thin homogeneous magnetized conducting layer of length $L$ and width $\ell$ contacted to an electric generator. The material used can be a conducting ferromagnet, or antiferromagnet, or altermagnet or any kind of materials that can be described with an effective magnetic field. We assume that the conducting layer is planar, invariant by translation along the $x$-axis $\ell \ll L$  (this excludes the region in contact with the power generator), and the two lateral edges are symmetric. 
 
We define the distribution of electric charge carriers by $n(y) = n_0 + \delta n(y)$, where $\delta n(y)$ is the charge accumulation and $n_0$ the homogeneous density in the electrically neutral system.  Due to the symmetry of the device and the global charge conservation we have $\int_{-\ell}^{+\ell} \delta n \dd{y} = 0$, and the total charge carrier density is constant $n_\text{tot} = \frac{1}{2 \ell}\int n \dd{y}$. For the sake of simplicity, we assume a global charge neutrality so that $n_\text{tot} = n_0$. On the other hand, the external generator injects a current along the $x$ direction, so that the global current flowing in the $x$ direction throughout the device is also constant along $x$, by definition of the galvanostatic condition. This is expressed as $\int_{- \ell}^{\ell} J_x(y) \dd{y} = 2 \ell J_x^0$. 

Furthermore, in Section V, we will consider the case where a Hall-current is injected into the load circuit contacted at the lateral edges. We then have to introduce the Joule dissipation due to the load resistance. Unlike for the $x$ direction, there is no external generator in the $y$ direction, so that the loop integral of current in the $\{yz\}$ plan has to be $0$ : $\oint_{ABCDA} J \dd{l}=0$, where the  integration path $ABCD$ is sketched in Fig.1b. This can be written equivalently as $\int_{AB} J \dd{l} + \int_{BCDA} J \dd{l} = \int_{- \ell}^{\ell} J_y(y) \dd{y} + \int_{BCDA} J \dd{l}=0$. The definite integral $\int_{BCDA} J \dd{l}$ is a constant that we do not know for the moment, and we will represent the constant as $\int_{BCDA} J \dd{l} = - 2l J_y^{0}$, where $J_y^{0}$ is determined in Section V. In summary, we can write the global constraint on the $y$ direction as $\int_{- \ell}^{\ell} J_y(y) \dd{y} = 2l J_y^{0}$.

\section{Joule dissipation including screening}
The charge accumulation is governed by the Poisson's equation $\nabla^2 V = -\frac{q}{\varepsilon} \delta n$, where $V$ is the electrostatic potential, $q$ is the electric charge, and $\varepsilon$ is the electric permittivity. The local electrochemical potential $\mu(x,y)$ - that takes into account not only the electrostatic potential $V$ but also the energy (or the entropy) responsible for the diffusion - is given by the expression \cite{DeGroot,Rubi}  (local equilibrium is assumed everywhere): 
\begin{equation}
  \mu = \frac{k T}{q} \ln \left( \frac{n}{n_0} \right) + V ,
  \label{mu}
\end{equation}
where $k$ is the Boltzmann constant and the temperature $T$ is the temperature of the heat bath in the case of a non-degenerate semiconductor, or the Fermi temperature $T_F$ in the case of a fully degenerate conductor. Poisson's equation now reads
\begin{equation}
  \nabla^2 \mu - \lambda_D^2 \frac{q}{\varepsilon} n_0 \nabla^2 \ln \left( \frac{n}{n_0} \right) + \frac{q}{\varepsilon} \delta n = 0,
  \label{poisson-mu}
\end{equation}
where $\lambda_D= \sqrt{\frac{k T \varepsilon}{q^2 n_0}}$ is the Debye-Fermi length ($T$ is the temperature and $k$ the Boltzmann constant). On the other hand, the transport equation under a magnetic field is given by Ohm's law:
\begin{equation}
\vec{J} = -\hat{\sigma} \vec{\nabla} \mu = - q n \hat{\eta} \vec{\nabla} \mu,
\label{Ohm}
\end{equation}
 where the transport coefficients are the conductivity tensor $\hat{\sigma}$ or the mobility tensor $\hat{\eta}$. 

We define the resistivity of the isotropic material $\rho$, the anisotropic magnetoresistance $\Delta \rho/\rho$, and the Hall resistivity $\rho_H$. Based on the symmetry property of the magnetic system and taking into account the Onsager reciprocity relations, the resistivity matrix takes the following form \cite{Seitz,Matrix}:
\begin{align}
\hat{\rho}=\left(\begin{array}{ccc}
    \rho+\Delta \rho m_{x}^{2} & \Delta \rho m_{x} m_{y}-\rho_{H} m_{z} & \Delta \rho m_{x} m_{z}+\rho_{H} m_{y} \\
    \Delta \rho m_{y} m_{x}+\rho_{H} m_{z} & \rho+\Delta \rho m_{y}^{2} & \Delta \rho m_{y} m_{z}-\rho_{H} m_{x} \\
    \Delta \rho m_{z} m_{x}-\rho_{H} m_{y} & \Delta \rho m_{z} m_{y}+\rho_{H} m_{x} & \rho+\Delta \rho m_{z}^{2}
  \end{array}\right)
  \label{Matrix}
\end{align}
where the unit vector $\vec m$ gives the direction of the magnetization with respect to the orthogonal basis attached to the Hall-bar. The unit vector $\vec m$ can be described as a function of the radial angle $\theta$ and the orthoradial angle $\varphi$, such that $m_x= sin (\theta) cos (\varphi), \, m_y= sin (\theta) sin (\varphi), \, m_z= cos \theta$.
In order to compare the Planar Hall effect with the Anomalous Hall effect, we choose the material of the Hall bar so that $\rho \gg \Delta \rho \gg \rho_H$ for the planar Hall effect (this is typically the case for $NiFe$ alloys \cite{Madon}) and $\rho  \gg \rho_H \gg \Delta \rho$ for the anomalous Hall effect (this is typically the case for $GdCo$ alloys \cite{Madon,ArXiv_2024}). In the following, the index $pl$ stands for PHE (left) and the index $an$ stands for AHE (right).  In addition, we only consider the behavior of the system on the $xy$-plane (translation invariance along $z$ axis) thus the matrix Eq.(\ref{Matrix}) can be reduced to a $2\times 2$ matrix:

\begin{align}
  \begin{aligned}[c]
    \hat{\rho}_{pl}=\left(\begin{array}{cc}
      \rho & \Delta \rho \, m_{x} m_{y} \\
      \Delta \rho \, m_{y} m_{x} & \rho \\
    \end{array}\right)
      \label{eq:Matrix}
  \end{aligned}
  \qquad \text{ and } \qquad
  \begin{aligned}[c]
    \hat{\rho}_{an}=\left(\begin{array}{cc}
      \rho & -\rho_{H} \, m_{z}  \\
      \rho_{H} \, m_{z}  & \rho \\
    \end{array}\right)
  \end{aligned}
\end{align}

We then define the ratio of the off-diagonal coefficient over the diagonal coefficient:
\begin{align}
  \begin{aligned}[c]
    \Theta_{pl} =   \frac{ \Delta \rho}{\rho} \, sin^2(\theta) \, sin(2 \varphi)
  \end{aligned}
  \qquad \text{ and } \qquad
  \begin{aligned}[c]
    \Theta_{an} = - \frac{\rho_H}{\rho} \, cos(\theta)
  \end{aligned}
  \label{Hall_Angle}
\end{align} 

The angular dependence of Eqs.(\ref{Hall_Angle}) appears directly in the Hall voltage (through the charge accumulation at the edges), and is an unambiguous signature that allows the AHE to be distinguished from PHE \cite{Madon,ArXiv_2024}. However, in our context (theory and experiment), we will choose the magnetization direction such that $\theta = \pi/2$ and $\varphi = - \pi/4$ for the PHE and $\theta = 0$ for AHE. If possible, we will also choose two materials such that $\Delta \rho / \rho \approx \rho_{H}/\rho $ in order to have $\Theta = \Theta_{an} = - \Theta_{pl}$. Furthermore, since $\Delta \rho / \rho$ and $\rho_{H}/\rho $ are of the order of 1\% maximum, we have $sin( \Theta ) \approx \Theta$ and we call $\Theta$ the ``Hall angle''.

Then the conductivity matrix follows as the inverse of resistivity matrix $\hat{\sigma}=\hat{\rho}^{-1}$ and the mobility tensor can be derived from $\hat{\sigma}=qn\hat{\eta}$.

\begin{align}
  \begin{aligned}[c]
    \hat{\eta}_{pl} =
    \begin{pmatrix}
       \eta    & \eta_{pl} \\
      \eta_{pl} & \eta
    \end{pmatrix}
    =  \eta
    \begin{pmatrix}
      1         &   - \Theta \\
 - \Theta & 1
    \end{pmatrix}
  \end{aligned}
  \qquad \text{ and } \qquad
  \begin{aligned}[c]
    \hat{\eta}_{an} =
    \begin{pmatrix}
      \eta    & \eta_{an} \\
      - \eta_{an} & \eta
    \end{pmatrix} = \eta
    \begin{pmatrix}
      1         & \Theta \\
      - \Theta & 1
    \end{pmatrix}
  \end{aligned}
\end{align}

 In the following, we want to express the joule dissipation density in terms of the current vector components $J_x$, $J_y$. The Joule dissipation power density is formally given by \begin{align}
  p_J = \vec J \cdot \vec \nabla \mu
\end{align} 

Using Ohm's law expressed in Eq.(\ref{Ohm}), the gradients of the chemical potentials $\vec \nabla \mu$ read
\begin{align}
  \begin{aligned}[c]
    \vec \nabla \mu^{pl}  &= -\dfrac{1}{qn} \hat \eta^{-1}_{pl} \vec J \\
  &= - \dfrac{1}{qn\eta (1   - \Theta^2)}  \begin{pmatrix}
    1         & \Theta \\
     \Theta & 1
  \end{pmatrix}  
  \begin{pmatrix} J_x \\ 
  J_y
  \end{pmatrix} \\
  \vec \nabla \mu^{pl} &= -\dfrac{n_0}{n \sigma_{pl}} 
   \begin{pmatrix} 
   J_x  + \Theta J_y \\
   \Theta J_x  + J_y 
   \end{pmatrix}
  \end{aligned}
  \qquad \text{ and } \qquad
  \begin{aligned}[c]
    \vec \nabla \mu^{an}   &= -\dfrac{1}{qn}  \hat \eta^{-1}_{an} \vec J  \\
  &= -\dfrac{1}{qn\eta (1 + \Theta^2)}
   \begin{pmatrix}
    1         & -\Theta \\
    \Theta & 1
  \end{pmatrix} 
   \begin{pmatrix} J_x  \\ 
  J_y
  \end{pmatrix} \\
  \vec \nabla \mu^{an} &= -\dfrac{n_0}{n \sigma_{an}}  
  \begin{pmatrix} 
  J_x  -\Theta J_y \\ 
  \Theta J_x  + J_y 
  \end{pmatrix}
  \end{aligned}
  \label{Ohm2}
\end{align}
where,  for convenience, we have introduced the two constant conductivities (i.e; the bulk conductivities for $y \gg \lambda_D$),:
\begin{align}
 \begin{aligned}[c]
  \sigma_{pl} &= qn_0 \eta (1 - \Theta^2)
  \end{aligned}
  \qquad \text{ and } \qquad
  \begin{aligned}[c]
\sigma_{an} &= qn_0 \eta (1 + \Theta^2)
  \end{aligned}
  \label{Conductivities}
\end{align}

Note that since $\rho$ is the intrinsic value of the resistivity of the material  \cite{Madon, ArXiv_2024} (that does not depend on $\Theta$) and assuming identical material for both AHE and PHE, we have  $\sigma_{pl} = \frac{1}{\rho(1 - \Theta^2)}$ and $\sigma_{ah} = \frac{1}{\rho(1 + \Theta^2)}$. This gives us
\begin{align}
  \begin{aligned}[c]
  p_J^{pl} &=  - \dfrac{n_0}{n \sigma_{pl}} \left(J_x^2 + J_y^2 + 2\Theta J_x J_y \right)
  \end{aligned}
  \qquad \text{ and } \qquad
  \begin{aligned}[c]
  p_J^{an} &=  - \dfrac{n_0}{n \sigma_{an}} \left(J_x^2 + J_y^2 \right)
  \end{aligned}
\end{align}

The expression of the Joule power dissipated by the system  reads:
\begin{align}
  P_J  &= S_\text{lat} \int_{-\ell}^{\ell}  p_J(y) \dd{y} 
\end{align} which gives respectively
\begin{align}
  \begin{aligned}[c]
    P_J^{pl} &= \frac{S_\text{lat}}{\sigma_{pl}} \int_{-\ell}^{\ell} \frac{n_0}{n} \left(J_x^2 + J_y^2 + 2\Theta J_x J_y \right) \dd{y}
  \end{aligned}
  \qquad \text{ and } \qquad
  \begin{aligned}[c]
    P_J^{an} &= \frac{S_\text{lat}}{\sigma_{pl}} \int_{-\ell}^{\ell} \frac{n_0}{n} \left(J_x^2 + J_y^2\right) \dd{y}
    \label{Joule}
  \end{aligned}
\end{align}
where $S_\text{lat}$ is the lateral surface of the Hall bar (product of the length $L$ by the thickness), and $2 \ell$ is the width. As expected the two expressions of the Joule heating for PHE and AHE are significantly different.

\section{Currents and charge densities without load circuit}

The stationary state is defined by the {\it least dissipation principle} that states that the current distributes itself so as to minimize Joule heating $P_J$ compatible with the constraints \cite{Onsager_Diss,Bruers,MinDiss}. Without load circuit, in the notation of Section III, the integration path ABCD is reduced to AB and the global constraint along the $y$ direction is zero: $\int_{- \ell}^{\ell} J_y(y) \dd{y} = 0$.

The global constraints read: 
 \begin{equation}
 \int_{- \ell}^{\ell} n(y) \dd{y} = 2 \ell n_0
\quad \textrm{and} \quad 
\int_{- \ell}^{\ell} J_x(y) \dd{y} = 2 \ell J_x^0
\quad \textrm{and} \quad 
 \int_{- \ell}^{\ell} J_y(y) \dd{y} = 0
\label{Constr}
 \end{equation}

We define for convenience the reduced power $\tilde{P_J} = \frac{q \eta (1\pm\Theta^2)}{S_\text{lat}}\, P_J$.
Let us introduce the two Lagrange multiplayers $\lambda_x$ and  $\lambda_n$ corresponding to the two constraints Eqs.(\ref{Constr}). The functional to be minimized then reads: 

\begin{align}
  \hspace{-1cm}
  \begin{aligned}[c]
    \tilde {\mathcal{P}}_J^{pl}  =   \int_{- \ell}^{\ell} \left ( \frac{J_x^2 + J_y^2 + 2\Theta J_x J_y}{n}  - \lambda_x \, J_x - \lambda_y \, J_y - \lambda_n \, n \right ) \dd{y}
  \end{aligned}
  \qquad \text{ and } \qquad
  \begin{aligned}[c]
    \tilde {\mathcal{P}}_J^{an}  =   \int_{- \ell}^{\ell} \left ( \frac{J_x^2 + J_y^2}{n} - \lambda_y \, J_y  - \lambda_x \, J_x - \lambda_n \, n \right ) \dd{y}
  \end{aligned}
\end{align}
The minimum corresponds to \begin{align}
  \frac{\funcd \tilde{\mathcal{P}}_J}{\funcd J_x} = 0 \quad \textrm{and} \quad  \frac{\funcd \tilde{\mathcal{P}}_J}{\funcd J_y} = 0  \quad \textrm{and} \quad  \frac{\funcd \tilde{\mathcal{P}}_J}{\funcd (n)} = 0  \label{Conds} 
\end{align}
which gives respectively 
\begin{align}
  \begin{aligned}[c]
      \begin{cases}
        2J_x + 2\Theta J_y = n \lambda_x \\
        2 J_y + 2\Theta J_x = n \lambda_y \\
        J_x^2 + J_y^2 + 2\Theta J_x J_y = - \lambda_n n^2
      \end{cases}
  \end{aligned}
  \qquad \text{ and } \qquad
  \begin{aligned}[c]
    \begin{cases}
      2J_x = n \lambda_x \\
      2 J_y = n \lambda_y \\
      J_x^2 + J_y^2 = - \lambda_n n^2
    \end{cases}
  \end{aligned}
\end{align}

Integrating the two first equations on both sides from $-l$ to $l$, and then applying the constraints, we can solve $\lambda_x$ and $\lambda_y$ explicitly. Thus, we can obtain $J_x$ and $J_y$ by re-inserting this result into the equations above. The solution then reads respectively:

\begin{align}
  \begin{aligned}[c]
      \begin{cases}
        J_x^{pl}(y) = J_x^0\dfrac{n(y)}{n_0}  \\
 J_y^{pl}(y) =   0
      \end{cases}
    \end{aligned}
    \qquad \text{ and } \qquad
    \begin{aligned}[c]
      \begin{cases}
        J_x^{an}(y) = J_x^0\dfrac{n(y)}{n_0}  \\
 J_y^{an}(y) = 0
    \end{cases}
  \end{aligned}
  \label{min}
\end{align}

As can be seen, the density of current is not homogeneous throughout the sample (since $n(y) \ne n_0$), and it has the same form for both AHE and PHE. 

Inserting the solution (\ref{min}) into the transport equations Eq.(\ref{Ohm2}), we deduce \begin{align}
 \begin{aligned}[c]
 \begin{cases}
        \partial_x \mu^{pl}(y) =  -\frac{J_x^0}{\sigma_{pl} }  \\
        \partial_y \mu^{pl}(y) =  - \frac{  \Theta J_x^0 }{\sigma_{pl} } 
      \end{cases}
    \end{aligned}
    \qquad \text{ and } \qquad
    \begin{aligned}[c]
      \begin{cases}
        \partial_x \mu^{an}(y) = -\frac{J_x^0}{\sigma_{an} } \\
        \partial_y \mu^{an}(y) = - \frac{ \Theta J_x^0}{\sigma_{an}  }
    \end{cases}
  \end{aligned}
\end{align} 

As a consequence, the electrochemical potential of the stationary state is harmonic in both cases: $\nabla^2 \mu = 0$. Since the profile of the lateral current $J_y(y)$ is defined by the charge density $n(y)$, the Poisson's equation Eq.(\ref{poisson-mu}) for $\nabla^2 \mu = 0$ gives the equation:
\begin{equation}
  \lambda_D^2 \partial_y^2 \ln \left(1 + \frac{\delta n }{n_0} \right) = \frac{\delta n}{n_0}.
  \label{poisson-final}
\end{equation} Assuming $\delta n \ll n_0$, we have, at the first order:
\begin{equation}
  \lambda_D^2 \partial_y^2 \left(\frac{\delta n }{n_0}\right) = \frac{\delta n}{n_0}.
\end{equation}  which gives us 
\begin{equation}
   \frac{\delta n(y)}{n_0}= A \ch(y/\lambda_D) + B\sh(y/\lambda_D)
   \label{Poisson}
\end{equation}

Once again, the boundary conditions for the density $n$ are not defined locally but globally by Eq.({\ref{Constr}}). We can specify the boundary conditions for $\partial_y\frac{\delta n(y)}{n_0}$ noting that for the stationary solutions of Eq.(\ref{min}) gives us for both systems \begin{align}
      E_y &= - \partial_y V \\ 
      &=  -\partial_y \mu +  \dfrac{kT}{q}  \partial_y  \left(\frac{\delta n }{n_0}\right) \\
      E_y &=  - \dfrac{\Theta J_x^0  }{\sigma_{h} } +  \dfrac{kT}{q}  \partial_y  \left(\frac{\delta n }{n_0}\right)
\end{align} 
where $\sigma_{h}$ is given by $\sigma_{pl}$ or $\sigma_{an}$.

Specifying these conditions at the edges of the sample for $E_y(\pm \ell) = E_y(\pm\infty)$, which correspond to the electric field at the exterior of the bar, assuming that the bar is surrounded by vacuum. If we define $2 \Delta^\pm E_\infty = E_y(\infty) 
\pm E_y(-\infty)$, then the approximated solution of $\delta n$ is given for both cases by:
\begin{align}
  \dfrac{q\lambda_D}{\varepsilon}\delta n(y)&= \Delta^- E_\infty  \dfrac{\ch(y/\lambda_D)}{\sh(l/\lambda_D)} +  \left(\Delta^+ E_\infty + \dfrac{  \Theta J_x^0 }{\sigma_{h} }   \right) \dfrac{\sh(y/\lambda_D)}{\ch(l/\lambda_D)}
  \label{eq:solutionsnolateral}
\end{align}  where $\sigma_{h}$ is here again given by $\sigma_{pl}$ or $\sigma_{an}$.

As can be seen, the the profiles of the screening are similar for both AHE and PHE in the perfect Hall-like device.  A difference can be seen (at the second order in $\Theta$) in terms of the conductivities $\sigma_{pl}$ and  $\sigma_{an}$ given in Eq.(\ref{Conductivities}). 

\section{Current Injection in the load circuit}
\label{sec:sections}

The solution found in the preceding section is valid as long as the dissipation due to charge leakage at the edges is negligible with respect to the dissipation inside the device. However, if it is no longer the case, the stationary regime should be reconsidered by introducing the dissipation due to the resistance of a lateral load circuit that connects the edges of the Hall bar.
In order to take into account this supplementary dissipation, we introduce the load resistivity $R_\ell$ ($\Omega \cdot \mathrm{m}^{ 2}$) of the lateral circuit. The power dissipated in the lateral circuit is:
\begin{equation*}
  P_\text{lat}  \ = \  \frac{S_\text{lat}\, \Delta \mu^2}{R_\ell}
  \label{Power_lat}
\end{equation*}
where $\Delta \mu = \mu (+ \ell) - \mu (- \ell)$ is the difference of the chemical potential between both edges. Note that due to our hypothesis of the invariance along $x$, we do not treat the case of a resistance contacted with a usual contact  to the two edges of the Hall bar. Indeed, such a contact would break the translation invariance symmetry along $x$, and would distort the current lines in a specific manner that depends on the details of the contact geometry and resistivity. Such a contact-specific effect is not related to the generic problem studied here. The relevant contact, as sketched in Fig. 1 (left), is an idealization for the ``perfect Hall bar'', but it can be approximated experimentally\cite{ArXiv_2024,Madon}. Incidentally, it is well known that the main advantage of the Corbino disk with respect to Hall-bar devices is precisely that it is much easier to design two quasi-perfect concentric equipotentials (circular symmetry) instead of two quasi-perfect longitudinal equipotentials (translational symmetry).

\subsection{Stationary Currents}
The difference of chemical potential can be expressed as a function of the current:
\begin{align}
  \begin{aligned}[c]
    \Delta \mu^{pl} \  &= 
    \  \int_{- \ell}^{+ \ell} \dd{y}\, \partial_y
    \mu \ \\ &=\  - \int_{- \ell}^{+ \ell} \dd{y}\,
    \frac{J_y + \Theta J_x}{qn \eta (1 - \Theta^2)}
  \end{aligned}
  \qquad \text{ and } \qquad
  \begin{aligned}[c]
    \Delta \mu^{an} \  &= 
    \  \int_{- \ell}^{+ \ell} \dd{y}\, \partial_y
    \mu \ \\ &=\  - \int_{- \ell}^{+ \ell} \dd{y}\,
    \frac{J_y + \Theta J_x}{qn \eta (1 + \Theta^2)}
\end{aligned}
\end{align} 

so that

\begin{align}
  \begin{aligned}[c]
    P_\text{lat}^{pl} \  = \  \frac{S_\text{lat}}{R_\ell(q \eta)^2  (1 - \Theta^2)^2}  \left( \int_{-
    \ell}^{+ \ell} \dd y \frac{J_y + \Theta J_x}{n} \right)^2
  \end{aligned}
  \qquad \text{ and } \qquad
  \begin{aligned}[c]
    P_\text{lat}^{an} \  = \  \frac{S_\text{lat}}{R_\ell(q \eta)^2  (1 + \Theta^2)^2}  \left( \int_{-
    \ell}^{+ \ell} \dd y \frac{J_y + \Theta J_x}{n} \right)^2
\end{aligned}
\label{PowerLeak}
\end{align}

As in the preceding section, we define the reduced power $\tilde{P} = \frac{q \eta (1 \pm \Theta^2)}{S_\text{lat}} P$. The total power dissipated is then:

\begin{align}
  \begin{aligned}[c]
    \tilde{P}^{pl} \  &= \  \tilde{P}_J + \tilde{P}_\text{lat} \\  
    &= \  \int_{- \ell}^{+ \ell} \dd
    y \frac{J_x^2 + J_y^2 + 2\Theta J_x J_y}{n} \  +
    \\  &\phantom{=} \alpha^{pl} \frac{2l}{n_0}  \left( \frac{n_0}{2 \ell} \int_{- \ell}^{+ \ell} 
    \dd y \frac{J_y + \Theta J_x}{n} \right)^2 
  \end{aligned}
  \qquad \text{ and } \qquad
  \begin{aligned}[c]
    \tilde{P}^{an} \  &= \  \tilde{P}_J + \tilde{P}_\text{lat} \\  
    &= \  \int_{- \ell}^{+ \ell} \dd
    y \frac{J_x^2 + J_y^2}{n} \  +
    \\ &\phantom{=}  \alpha^{an} \frac{2l}{n_0}  \left( \frac{n_0}{2 \ell} \int_{- \ell}^{+ \ell} 
    \dd y \frac{J_y + \Theta J_x}{n} \right)^2 
  \end{aligned}
  \label{Power}
\end{align} 

where we have introduced the {\bf dimensionless control parameter} $\alpha^{h} = \frac{2 \ell } {R_\ell \sigma_{h}}$. Note that this control parameter $\alpha$ is defined by the ratio $\alpha = \frac{R }{R_\ell}$ of the resistance of the material $R \equiv \frac{V}{J_x^0} = \frac{2 \ell}{\sigma_{h}}$ over the load resistance $R_\ell$ \cite{ArXiv_2024}. The case $\alpha \rightarrow 0$ corresponds to perfect Hall bar while the case $\alpha \rightarrow \infty$ correspond to the perfect Corbino disk \cite{Benda,Madon}. The exponent of index $h$ denotes respectively $h = pl$ for the planar Hall system and $h = an$ for the anomalous Hall system. We define for convenience the constant $A^{h}$:
\begin{align}
    A^{h} \equiv  \frac{n_0}{2 \ell} \int_{\ell}^{\ell}\frac{ \Theta J_x + J_y }{n}\dd{y}.
  \label{A}
\end{align} 
At this stage, we can adopt the constraint on $J_y(y)$ as discussed in the section above:
\begin{align} \int_{- \ell}^{\ell} J_y(y) \dd{y} = 2 \ell J_y^{0, h} \label{ConstrJY0},
\end{align}
where $J_y^{0, h}$ is some constant for each system which will be determined later.

The minimization of the corresponding functional $\tilde{\mathcal{P}}$ now reads:
\begin{align}
  \begin{aligned}[c]
  \alpha^{pl} \, A^{pl} \, \Theta + J_x + \Theta J_y = \frac{\lambda_x n}{2},
  \\
  \alpha^{pl} \, A^{pl} \,  + J_y + \Theta J_x = \frac{\lambda_y n}{2},
  \end{aligned}
  \qquad \text{ and } \qquad
  \begin{aligned}[c]
  \alpha^{an} \, A^{an} \, \Theta + J_x = \frac{\lambda_x n}{2},
  \\
    \alpha^{an} \, A^{an} \,  + J_y  = \frac{\lambda_y n}{2}.
  \end{aligned}
\end{align}

Once again, applying the global constraints given by Eqs.(\ref{Constr}) along with the third replaced by Eq. (\ref{ConstrJY0}), we can immediately solve from these two equations 
\begin{align}
  \begin{aligned}[c]
  \lambda_x &= \frac{2}{n_0} \left(\alpha^{pl} A^{pl} \Theta + J_x^0 + \Theta J_y^{0, pl}\right) \\
  \lambda_y &= \frac{2}{n_0} \left(\alpha^{pl} A^{pl} + \Theta J_x^0 + J_y^{0, pl}\right)
  \end{aligned}
  \qquad \text{ and } \qquad
  \begin{aligned}[c]
    \lambda_x &= \frac{2}{n_0} \left(\alpha^{an} A^{an} \Theta + J_x^0\right) \\
    \lambda_y &= \frac{2}{n_0} \left(\alpha^{an} A^{an} + J_y^{0, an}\right)
  \end{aligned}
\end{align} 
Then it follows that:
\begin{align}
  \begin{aligned}[c]
    J_x &= \frac{n}{n_0}J_x^0 \\
    J_y &= \frac{n}{n_0}J_y^{0, pl} + \alpha^{pl} A^{pl} \left( \frac{n}{n_0} -1\right)
  \end{aligned}
  \qquad \text{ and } \qquad
  \begin{aligned}[c]
    J_x &= \frac{n}{n_0}J_x^0 + \alpha^{an} A^{an} \Theta \left( \frac{n}{n_0} - 1\right) \\
    J_y &= \frac{n}{n_0}J_y^{0, an} + \alpha^{an} A^{an} \left( \frac{n}{n_0} -1\right)
  \end{aligned}
\end{align}

For the sake of simplicity, let us focus on a system which obeys the local conservation of electric charges: $\mathbf{\nabla} \cdot \mathbf{J}=0$, which leads to $\frac{\partial J_x}{\partial x}+\frac{\partial J_y}{\partial y}=0$. Since the system is invariant along the $x$ direction, so $\frac{\partial J_x}{\partial x}=0$. Then this will lead to the result that $\frac{\partial J_y}{\partial y}=0$. Thus, this requires that coefficient in factor of $n(y)$ vanishes, which leads to $J_y^{0}=-\alpha A$. We get eventually:

\begin{align}
  \begin{aligned}[c]
    J_x &= \frac{n}{n_0}J_x^0 \\
    J_y &=  J_y^{0, pl}
  \end{aligned}
  \qquad \text{ and } \qquad
  \begin{aligned}[c]
    J_x &= \frac{n}{n_0}J_x^0 - J_y^{0, an} \Theta \left( \frac{n}{n_0} - 1\right) \\
    J_y &=  J_y^{0, an}
  \end{aligned}
  \label{mincc}
\end{align}

  Re-inserting Eq.(\ref{mincc}) into the definition of $A^{pl/an}$  Eq.(\ref{A}), we obtain the following equation
  \begin{align}
    \begin{aligned}[c]
    J_y^{0, pl} \, \left(1 + \alpha \int \frac{n_0}{n}\frac{\dd{y}}{2\ell}\right)&= -\alpha \Theta J_x^0  
    \end{aligned}
    \qquad \text{ and } \qquad
    \begin{aligned}[c]
      J_y^{0, an} \, \left(1 + \alpha \Theta^2  + \alpha(1- \Theta^2) \int \frac{n_0}{n}\frac{\dd{y}}{2\ell}\right)&= - \alpha \Theta J_x^0  
    \end{aligned}
  \end{align}

  Assuming that $\delta n \ll n_0$, we have {\it at the first order} $ \frac{n_0}{2\ell}\int_{-\ell}^{+ \ell} \frac{\dd{y}}{n(y)} \simeq 1$, thus we can solve 
  \begin{align}
    \begin{aligned}[c]
      J_x &= \frac{n}{n_0}J_x^0 \\
      J_y &= - \frac{\alpha }{1+\alpha}\Theta J_x^0
    \end{aligned}
    \qquad \text{ and } \qquad
    \begin{aligned}[c]
      J_x &= \frac{n}{n_0}J_x^0 + \Theta^2\frac{\alpha }{1  +\alpha}\left( \frac{n}{n_0} - 1\right) J_x^0 \\
      J_y &= - \frac{\alpha }{1  +\alpha}\Theta J_x^0
  \end{aligned}
  \label{eq:mincircuit}
\end{align} 
  
The results Eqs.(\ref{eq:mincircuit}) show that, at the first order in the charge accumulation, the distributions of the transverse Hall-current $J_y$ are identical for the both the AHE and the PHE, and a difference can be seen at the second order in the Hall angle $\Theta$ for the longitudinal current $J_x$.
The well-known expressions are recovered for the extreme cases, that are the perfect Hall bar and the Corbino disk. For an ideal Hall bar, the load resistance is infinite, and this implies $\alpha =0$. We thus see that Eq.(\ref{eq:mincircuit}) converges to Eq.(\ref{min}) as $\alpha \to 0$.  

Conversely, in a Corbino disk, we have $\alpha \rightarrow +\infty$ and the solutions converge to the well-known result
 \begin{align}
    J_x^{Cor} = \frac{n}{n_0}J_x^0 \quad \text{and} \quad
    J_y^{Cor} =  -\Theta J_x^0
  \end{align} 
for both PHE and AHE. Once again, we see that the second law of thermodynamics levels out the fundamental differences that define the two processes - PHE vs. AHE - at the microscopic scales. 

\subsection{Power Injected in the load circuit}

The total power $\tilde P = \tilde P_J + \tilde P_\text{lat}$ - given in Eq.(\ref{Power}) - is the sum of the Joule heating $\tilde P_J$ dissipated inside the Hall device, and the power $\tilde P_\text{lat}$ dissipated into the lateral passive circuit. Inserting the stationary state Eqs.(\ref{eq:mincircuit}) and using the first global condition in Eqs.(\ref{Constr}) (assuming that $\delta n \ll n_0$, we have $\int_{-\ell}^{+ \ell} \frac{\dd{y}}{n(y)} \simeq 2 \ell/n_0$), the power dissipated in the lateral circuit reads:
\begin{align}
  \begin{aligned}[c]
    P^{pl}_\text{lat}(\alpha)  &\simeq  \frac{2 \ell S_\text{lat}(J_x^0)^2} {\sigma_{pl}} \dfrac{\alpha}{(1+\alpha)^2}\, \Theta^2
  \end{aligned}
\, \, \,   \text{ and } \, \, \, \,  \, \,
  \begin{aligned}[c]
    P^{an}_\text{lat}(\alpha)  &\simeq  \frac{2 \ell S_\text{lat}(J_x^0)^2} {\sigma_{an}} \dfrac{\alpha}{(1+\alpha)^2} \, \Theta^2  \left(1 + 2  \Theta^2 \alpha \left( \frac{n}{n_0} - 1\right) \right)
\end{aligned}
\label{Power_Approx1} 
\end{align} 

 \begin{figure}[ht]
          \centering
          \includegraphics[width=0.48\textwidth]{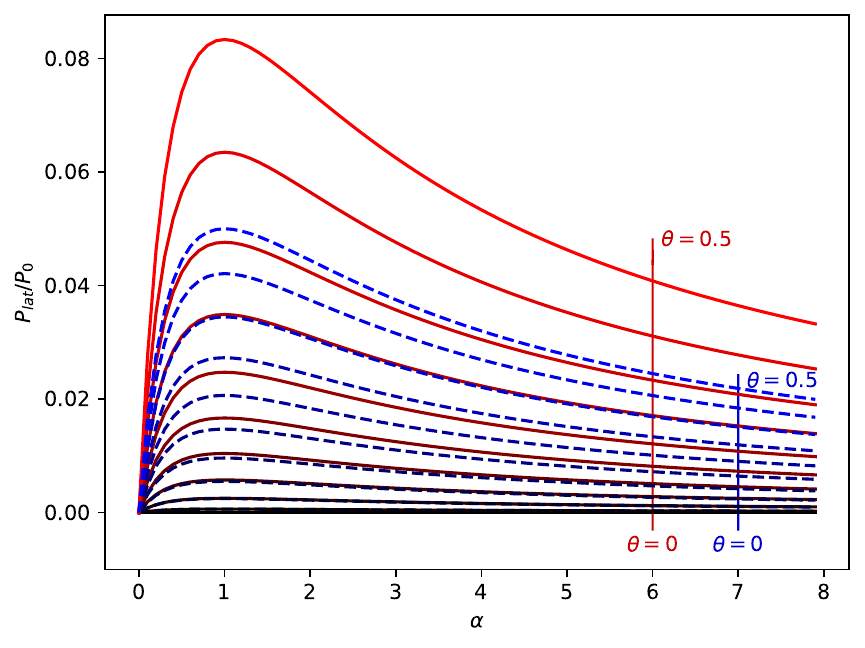}
            \includegraphics[width=0.48\textwidth]{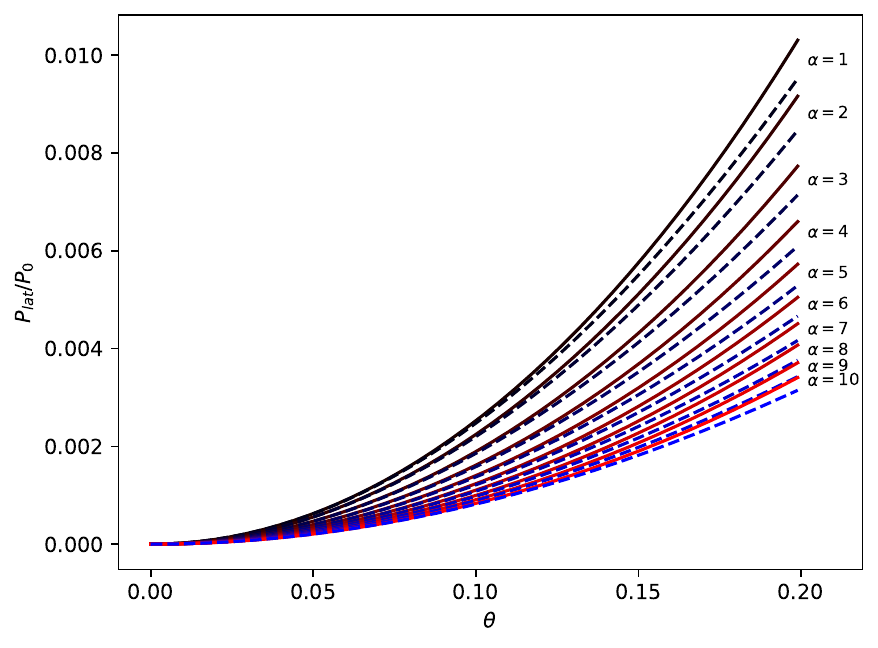}
         \caption{Power efficiency $P_{lat}/P_0$ for both AHE (red lines) and PHE (blue dashed line). Left:  Efficiency as a function of the ratio $\alpha = R/R_{\ell}$ for values of $\Theta$ ranging from 0 to 0.5 with increments of 0.05. Right: Efficiency as a function of the Hall angle $\Theta$. }
 \end{figure}

When we consider the power dissipated in the material in the absence of the Hall-like effect (this can be performed experimentally by adjusting the magnetization state according to Eq.(\ref{Hall_Angle}))
 \begin{align}
  P_0 \equiv P_J(\Theta = 0) = \dfrac{2 \ell S_\text{lat}(J_x^0)^2}{q \eta n_0}
  \label{Ref_Power}
\end{align} 
This leads to a very simple expression of  the power efficiency due to the injection of the Hall-current:
\begin{align}
  \begin{aligned}[c]
   \frac{ P^{pl}_\text{lat}}{P_0}  &\simeq   \dfrac{\alpha}{(1+\alpha)^2} \frac{\Theta^2} {1 - \Theta^2} 
  \end{aligned}
  \qquad \text{ and } \qquad
  \begin{aligned}[c]
    \frac{P^{an}_\text{lat}}{P_0}  &\simeq  \dfrac{\alpha}{(1+\alpha)^2} \frac{\Theta^2} {1 + \Theta^2}   \left(1 + 2  \Theta^2 \alpha \left( \frac{n}{n_0} - 1\right) \right) \\
    &\simeq   \dfrac{\alpha}{(1+\alpha)^2} \frac{\Theta^2} {1 + \Theta^2}  
\end{aligned}
\label{Power_Results}
\end{align}

The function $\alpha \mapsto \frac{\alpha}{(1+\alpha)^2}$ reaches a maximum at  $\alpha = 1$ decreases to $0$ at $\alpha = +\infty$. The two expressions $P^{an}_\text{lat}$ and  $P^{pl}_\text{lat}$ coincide at the lowest order in $\Theta$.

Equations (\ref{Power_Results}) shows that power efficiency for both effects is the same as a function of $\alpha$, but the dependence as a function of $\Theta$ differs. This behavior can be observed in Fig.2 left  panel (as a function of the variable $\alpha$) and Fig.2 right panel (as a function of the variable $\Theta$) for AHE (red continuous lines) and PHE (blue dashed lines). Particularly, both curves have a unique maximum for the resistance matching $\alpha = 1$ (i.e., $R= R_{\ell}$) between the two sub-circuits. In other terms, the {\it maximum transfer theorem} is verified in both Hall-like systems. But this result does not implies that the Kirchhoff's circuit law is valid. Indeed, this is not the case, since both stationary states are different while the lumped-element circuits for AHE and PHE would be identical (see Fig.1).

Besides, in standard materials, Anomalous-Hall angles and planar-Hall angles are small (of the order of 1\% in usual materials), so that the difference of power efficiency between AHE and PHE as a function of the Hall angle is small, but easy to access experimentally (incidentally, the difference of dissipated power between the Hall-bar and the Corbino disk is also at the second order in $\Theta_{an}$). Note that if - instead of Eq.(\ref{Ref_Power}) - we define the reference power as $P_0^{pl,an}(\alpha = 0)= 2 \ell S_{latt} \, (J_x^0 )^2/\sigma_{pl,an}$ (i.e. the power of the perfect Hall bar for AHE and PHE separately), there is no difference between the two power efficiencies.
 
 \subsection{Charge accumulation}

It is now necessary to calculate the expression of the charge accumulation $\delta n(y)$ at the edges of a Hall bar contacted to the load circuit. The simple result obtained in section III is modified.

Assuming the solutions given by Eqs.(\ref{eq:mincircuit}), we rewrite $\vec \nabla \mu $ given by Eq.(\ref{Ohm2}) as  

\begin{align}
  \begin{aligned}[c]
    &\begin{cases}
      \partial_x \mu^{pl} =  -\dfrac{n_0}{n \sigma_{pl}} (J_x  + \Theta J_y)  \\
      \partial_y \mu^{pl} = -\dfrac{n_0}{n \sigma_{pl}} (\Theta J_x  + J_y)
    \end{cases}\\
    &\begin{cases}
      \partial_x \mu^{pl} =  -\dfrac{J_x^0}{ \sigma_{pl}} (1  -\frac{n_0}{n}\frac{\alpha }{1+\alpha}\Theta^2)  \\
      \partial_y \mu^{pl} = - \dfrac{J_x^0\Theta }{\sigma_{pl}} (1 - \frac{n_0}{n}\frac{\alpha }{1+\alpha})
    \end{cases}
  \end{aligned}
  \qquad \text{ and } \qquad
  \begin{aligned}[c]
    &\begin{cases}
      \partial_x \mu^{an} =  -\dfrac{n_0}{n \sigma_{an}} (J_x  -\Theta J_y)  \\
      \partial_y \mu^{pl} = -\dfrac{n_0}{n \sigma_{an}} (\Theta J_x  + J_y)
    \end{cases} \\
    &\begin{cases}
      \partial_x \mu^{an} =  -\dfrac{J_x^0}{\sigma_{an}} ( 1 + \Theta^2\frac{\alpha }{1  +\alpha})  \\
      \partial_y \mu^{pl} = -\dfrac{J_x^0 \Theta}{\sigma_{an}} ( 1 + \Theta^2\frac{\alpha }{1  +\alpha} - \frac{n_0}{n}\frac{\alpha }{1  +\alpha}\left( 1 + \Theta^2 \right)  )
    \end{cases} 
  \end{aligned}
\end{align}

We get 
\begin{align}
  \begin{aligned}[c]
    \nabla^2 \mu^{pl} &= - \dfrac{J_x^0\Theta }{\sigma_{pl}} \dfrac{\alpha }{1+\alpha} \dfrac{\partial_y(n/n_0)}{(n/n_0)^2}\\
    &= - \dfrac{J_x^0 }{qn_0 \eta }\dfrac{\Theta }{1 - \Theta^2} \dfrac{\alpha }{1+\alpha} \dfrac{\partial_y(n/n_0)}{(n/n_0)^2}\\
    &= - \dfrac{q \lambda_D}{\varepsilon} n_{pl}  \dfrac{\partial_y(n/n_0)}{(n/n_0)^2}
  \end{aligned}
  \qquad \text{ and } \qquad
  \begin{aligned}[c]
    \nabla^2 \mu^{an} &= - \dfrac{J_x^0\Theta (1+\Theta^2)}{\sigma_{an}} \dfrac{\alpha }{1+\alpha} \dfrac{\partial_y(n/n_0)}{(n/n_0)^2}\\ 
    &= - \dfrac{J_x^0}{qn_0 \eta}\Theta  \dfrac{\alpha }{1+\alpha} \dfrac{\partial_y(n/n_0)}{(n/n_0)^2}\\ 
    &= - \dfrac{q \lambda_D}{\varepsilon} n_{an} \dfrac{\partial_y(n/n_0)}{(n/n_0)^2}
  \end{aligned}
\end{align} 

The density of carriers reads:
\begin{align}
  \begin{aligned}[c]
    n_{pl} &=  \dfrac{\varepsilon} {q \lambda_D} \dfrac{J_x^0 }{qn_0 \eta} \dfrac{\Theta }{1 - \Theta^2}\dfrac{\alpha }{1+\alpha} \\
    &= \nu \dfrac{\Theta }{1 - \Theta^2} \dfrac{\alpha }{1+\alpha} 
  \end{aligned}
  \qquad \text{ and } \qquad
  \begin{aligned}[c]
    n_{an} &= \dfrac{\varepsilon} {q \lambda_D} \dfrac{J_x^0}{qn_0 \eta} \Theta\dfrac{\alpha }{1+\alpha}\\
    &= \nu \,\Theta \dfrac{\alpha }{1+\alpha} 
  \end{aligned}
\end{align}
where we have set $\nu =  \varepsilon \, J_x^0 / (q^2 \lambda_D  n_0 \eta ) $. In this case, we find that the chemical potential is not harmonic. The resulting formulation of Poisson's equation Eq.(\ref{poisson-mu}) is thus the same for both system and is given by 
\begin{align}
   - \lambda_D^2  n_0 \nabla^2 \ln \left(1 + \frac{\delta n}{n_0} \right) - \lambda_D  n_{h}\dfrac{\partial_y(\delta n/n_0)}{(n/n_0)^2}  +  \delta n = 0
\end{align} where we recall that the index $h$ in $n_h$ is given by $h=pl$ for the planar Hall case and by $h=an$ for the anomalous Hall case. We can linearize the equations in the regime $\delta n / n_0 \ll 1$ to get
\begin{align}
    -  \lambda_D^2 \partial_y^2 \left(\frac{\delta n }{n_0}\right) - \frac{n_{h}}{n_0}\lambda_D \partial_y \left(\frac{\delta n }{n_0}\right)  + \frac{\delta n }{n_0} = 0
\end{align}  which can be solved exactly to give in both cases 
\begin{align}
  \begin{aligned}[c]
    \dfrac{q\lambda_D}{\varepsilon} \delta n (y) &= e^{\frac{r_1 y}{\lambda_D}} \dfrac{E_{+\infty} e^{- \frac{r_2 \ell}{\lambda_D}} - E_{-\infty} e^{\frac{r_2 \ell}{\lambda_D}} + 2\sh(\frac{r_2 \ell}{\lambda_D}) \frac{J_x^0 \Theta}{\sigma_{pl}(1+\alpha)}  }{2 r_1 \sh((r_1 - r_2) \frac{\ell}{\lambda_D}) } + e^{\frac{r_2 y}{\lambda_D}}\dfrac{E_{+\infty} e^{- \frac{r_1 \ell}{\lambda_D}} - E_{-\infty} e^{\frac{r_1\ell}{\lambda_D}} + 2\sh(\frac{r_1 \ell}{\lambda_D}) \frac{J_x^0\Theta}{\sigma_{pl}(1+\alpha)}  }{2 r_2 \sh((r_2 - r_1) \frac{\ell}{\lambda_D}) }
  \end{aligned}
\end{align} with 
\begin{align} 
\begin{aligned}[c]
  r_1^{pl} &= \dfrac{1}{2}\left(-\dfrac{\nu}{n_0} \dfrac{\Theta }{1 - \Theta^2} \dfrac{\alpha }{1+\alpha}  - \sqrt{\left(\dfrac{\nu}{n_0}\dfrac{\Theta }{1 - \Theta^2} \dfrac{\alpha }{1+\alpha}\right)^2 + 4}\right)\\
    r_2^{pl} &= \dfrac{1}{2}\left(-\dfrac{\nu}{n_0} \dfrac{\Theta }{1 - \Theta^2}\dfrac{\alpha }{1+\alpha} + \sqrt{\left(\dfrac{\nu}{n_0}\dfrac{\Theta }{1 - \Theta^2}\dfrac{\alpha }{1+\alpha}\right)^2 + 4}\right)
  \end{aligned}
  \text{ and }
  \begin{aligned}[c]
    r_1^{an} &= \dfrac{1}{2}\left(-\dfrac{\nu}{n_0}\Theta\dfrac{\alpha }{1+\alpha} - \sqrt{\left(\dfrac{\nu}{n_0}\Theta\dfrac{\alpha }{1+\alpha}\right)^2 + 4}\right)\\
    r_2^{an} &= \dfrac{1}{2}\left(-\dfrac{\nu}{n_0}\Theta\dfrac{\alpha }{1+\alpha} + \sqrt{\left(\dfrac{\nu}{n_0}\Theta\dfrac{\alpha }{1+\alpha}\right)^2 + 4}\right)
  \end{aligned}
  \label{Parametres_r}
\end{align} 

\begin{figure}[ht]
          \centering
          \includegraphics[width=0.48\textwidth]{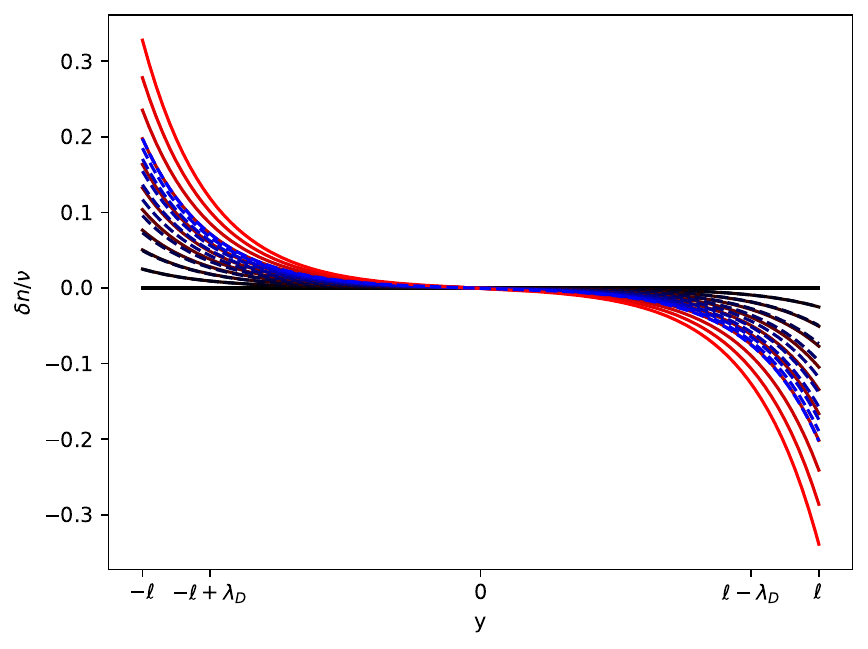}
            \includegraphics[width=0.48\textwidth]{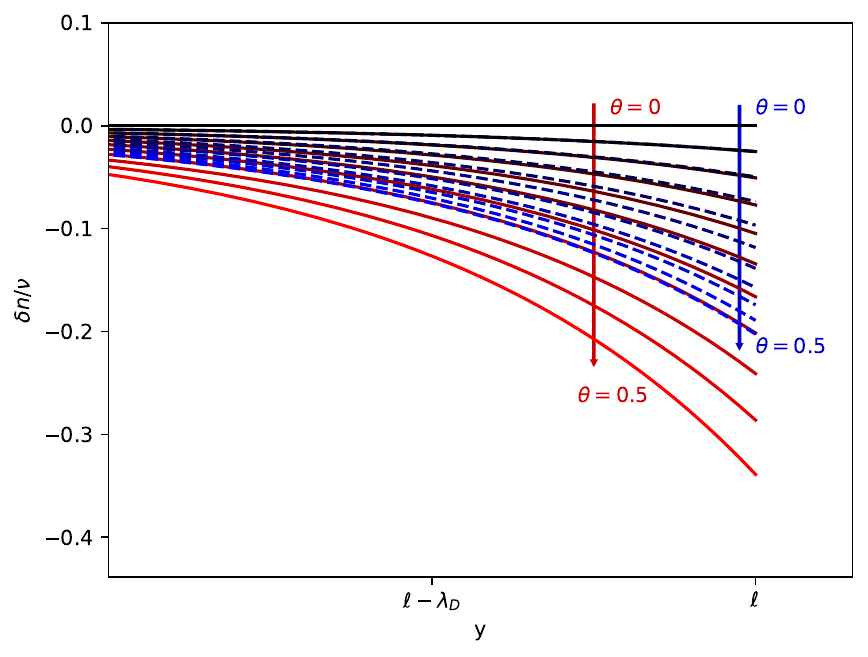}
         \caption{The profiles of the charge accumulations $\delta n(y)/\nu$ for both AHE (blue) and PHA (red) for different values of the Hall angle $\Theta$ at the maximum $\alpha =1$  (i.e., $R=R_{\ell}$) and for $\nu = 0.1n_0$. Left: over the whole sample.  Right: close to the right boundary. In both panels, the angle $\Theta$ range from $0$ to $0.5$ by increments of $0.05$. }
         \label{fig:profile}
 \end{figure}

Without other external source of electric fields, we have 
\begin{align}
    \dfrac{\delta n}{n_0}(y) &=\dfrac{\nu}{n_0}\dfrac{\Theta }{1 \pm \Theta^2}\dfrac{1 }{1+\alpha} \dfrac{1}{\sh((r_2 - r_1) \frac{\ell}{\lambda_D})}  \left(\sh(\frac{r_1 \ell}{\lambda_D}) \dfrac{ e^{\frac{r_2 y}{\lambda_D}} }{r_2 } - \sh(\frac{r_2 \ell}{\lambda_D}) \dfrac{ e^{\frac{r_1 y}{\lambda_D}} }{r_1  } \right) 
  \label{Densities}
\end{align} which, in the limit $\alpha \rightarrow 0$, adequately gives us back the distribution of Eq.(\ref{eq:solutionsnolateral}). The profiles of the charge accumulations $n(y)$ are functions of the parameters $\lambda_D/\ell$, $\nu$, $\alpha$ and $\Theta$. Despite the complexity of the expressions Eqs(\ref{Parametres_r}) and Eqs.(\ref{Densities}), the only difference between PHE and AHE is due to the difference as a function of $\Theta$ (see Fig.\ref{fig:profile}). We can see that changing the value of the Hall angle $\Theta$ will only change the amplitude of the distribution.

 \section{Conclusion}
In the context of the study of Hall currents injected into a lateral load circuit (Fig.1), we have compared the stationary states of the planar Hall effect (PHE)  with that of the anomalous Hall effect (AHE) - all other parameters being equivalent. The calculation is based on the principle of minimum power dissipation under the global constraints imposed to the system, and taking into account the screening effect. We derived analytical expressions of the distribution of charge carriers and currents for arbitrary values of the load resistance and Hall angle, at the first order in the charge accumulation $\delta n/n_0$. The well-known limiting cases corresponding to the perfect Hall bar (infinite load resistance) and the Corbino disk (zero load resistance) are recovered. \\

Despite the fundamental difference between the two effects at the microscopic scales - namely time-reversal-symmetry breaking - the accumulation of charge carriers, the distribution of currents, and the power efficiency of the injected Hall currents are surprisingly similar for both AHE and PHE (after having fixed the direction of the magnetization $\vec m$, according to Eqs.(\ref{Hall_Angle})). The difference is then due to the Hall-angle dependence of the conductivities (Eq.(\ref{Conductivities})), i.e. observable at the second order in $\Theta$ ($\Theta$ is of the order of 1\% in ordinary materials). In the case of the power efficiency, both curves have a maximum that corresponds to the resistance matching between the load circuit and the Hall bar. In other terms, {\it the maximum transfer theorem} is satisfied for both AHE and PHE. This property is not trivial because the system (shown in Fig.1) cannot be reduced to a simple lumped-element circuit. These results are an illustration of the importance of the second law of thermodynamics for dissipative transport processes, that levels out the peculiarities found at the microscopic scales, and that invalidates the prediction of dissipationless transport for macroscopic Hall-like systems.\\

It is worth noting that the results found are valid whatever the mechanisms responsible for the AHE: either due to spin-orbit scattering, topological properties of the band structure (Berry connection), as long as it leads to an effective magnetic field. This is also true for the PHE: it is valid whatever the mechanisms leading to axial anisotropy. Yet, the main limitations of the model are the quasi-classical approach and the local equilibrium hypothesis. The case of quantum anomalous Hall effect or unconventional anomalous Hall effect could be beyond the limit of validity of this model.  However, the specificity of nanoscopic or quantum characteristics would now be easily determined and quantified with a comparison of the measurements performed on these systems with that presented here.


\section*{References}
\begin{thebibliography}{0}

\bibitem{Carnot} Sadi Carnot, {\it R\'eflexion sur la puissance motrice du feu et sur les machines propres \`a d\'evelopper cette puissance}, 
Bachelier Libraire, Paris 1824 and Sadi Carnot and Robert Fox: {\it Reflexions on the Motive Power of Fire: A Critical Edition with the Surviving Scientific Manuscripts}, Manchester University Press (1986).

 \bibitem{Hall} E. H. Hall, {\it On a new action of the magnet on electric currents}, Am. J. Math. \textbf {2}, 287 (1879).
  
 \bibitem{Onsager1} L. Onsager, {\it Reciprocal relations in irreversible processes I}, Phys. Rev. {\bf 37}, 405 (1931)
  
  \bibitem{Onsager2} L. Onsager, {\it Reciprocal relations in irreversible processes II}, Phys. Rev. {\bf 38}, 2265 (1931)
  
\bibitem{Nobel_Lecture} The Nobel Prize in Chemistry 1968 was awarded to Lars Onsager ``for the discovery of the reciprocal relations bearing his name, which are fundamental for the thermodynamics of irreversible processes" https://www.nobelprize.org/prizes/chemistry/1968/summary/ 

\bibitem{Luttinger} R. Karplus and J. M. Luttinger, {\it Hall Effect in Ferromagnetics}, Phys. Rev. {\bf 95} 1154 (1954). 

\bibitem{Kondo} J. Kondo, {\it Anomalous Hall Effect and Magnetoresistance of Ferromagnetic Metals} Prog. Theo. Phys. {\bf 27} 772 (1962).

\bibitem{Berger} L. Berger, {\it Side-Jump Mechanism for the Hall Effect in Ferromagnets} Phys. Rev. B {\bf 2} 4559 (1970).

\bibitem{Nozieres} Ph. Nozi\`eres and C. Lewiner, {\it A simple theory of the anomalous Hall effect in semiconductors}, J. de Physique {\bf 34} (1973) 901.

\bibitem{Bruno} A. Crepieux and P. Bruno, {\it  Theory of anomalous Hall effect from the Kubo formula and the Dirac equation}, Phys. Rev. B {\bf 64}, (2001) 014416.

\bibitem{Haldane} F. D. M. Haldane, {\it Berry Curvature on the Fermi Surface: Anomalous Hall Effect as a Topological Fermi-Liquid Property}, Phys. Rev. Lett. {\bf 93}, 206602 (2004) 

  \bibitem{AHE} N. Nagasoa et al. {\it Anomalous Hall effect}, Rev. Mod. Phys. {\bf 82}, 1539 (2010).
  
  \bibitem{Altermag} Libor Smerjkal, Jairo Sinova, and Tomas Jungwirth, {\it Emerging Research Landscape of Altermagnetism}, Phys. Rev. X {\bf 12}, 04002 (2022).

\bibitem{Mazin} Igor Mazin, {\it Altermagnetism Then and Now}, Physics 17, 4 (2024) DOI: 10.1103/Physics.17.4


\bibitem{Dyakonov} M. I. Dyakonov and V. I. Perel, {\it Possibility of orienting electron spins with current}, Sov. Phys. JETP Lett. 13, 467 (1971).
\bibitem{Onoda} M. Onoda, S. Murakami, and N. Nagaosa, {\it Hall Effect of Light}, Phys. Rev. Lett. 93, 083901 (2004).


 \bibitem{PHE_Review} Amir Elzawy et al. {\it Current trends in planar Hall effect sensors: evolution, optimization, and applications}, J. Phys. D: Appl. Phys. {\bf 54} 353002 (2021).
    
 \bibitem{Schuhl} A. Schuhl, F. Nguyen Van Dau and J. R. Childress, {\it Low-field magnetic sensors based on the planar Hall effect} Appl. Phys. Lett. {\bf 66}, 2751 (1995).
     https://doi.org/10.1063/1.113697    

 \bibitem{Krivorotov} Christopher Sanfranski, Eric A. Montoya and Ilya N. Krivorotov {\it Spin-orbit torque driven by a planar Hall current},  Nature Nanotechnology, https://doi.org/10.1038/s41565-018-0282-0
    
\bibitem{SOT} Oiming Shao et al. {\it Roadmap of Spin Orbit Torques}, IEEE Trans. Mag. {\bf 57} (2021),

  \bibitem{Benda} R. Benda, E. Olive, M. J. Rub\`i and J.-E. Wegrowe {\it Towards Joule heating optimization in Hall devices}, Phys. Rev. B  {\bf 98}, 085417 (2018).
    
  \bibitem{Popovic} R.S. Popovic, {\it Hall Effect Devices}, IoP Publishing, Bristol and Philadelphia 2004 (Second Edition): Chapter 4, Paragraph 4.4: {\it The Hall current mode of operation}.
  
   \bibitem{Putley} E. H. Putley, The Hall Effect and Semiconductor Physics (Dover, New York 1968).

  
  \bibitem{Moelter} Moelter et al. Electric potential in the classical Hall effect: An unusual boundary-value problem  Am. J. Phys {\bf 66}, 668 (1998) https://doi.org/10.1119/1.18931

 

  \bibitem{Gyrator}   Giovanni Viola and David P. DiVincenzo,  Hall Effect Gyrators and Circulators , Phys. Rev. X {\bf 4}, 021019 (2014)
R. F. Wick  Solution of the Field Problem of the Germanium Gyrator , J. Appl. Phys. 25, 741
 

  
  \bibitem{JAP1}  M. Creff, F. Faisant, M. Rub\`i, J.-E. Wegrowe {\it surface current in Hall devices}, J. Appl. Phys. {\bf 128},  054501 (2020). 
  https://doi.org/10.1063/5.0013182.
  
  \bibitem{JAP2}   P.-M. D\'ejardin and J-E. Wegrowe {\it Stochastic description of the stationary Hall effect}, J. Appl. Phys. {\bf 128},  184504 (2020)
  
  \bibitem{JAP3} F. Faisant, M. Creff, J.-E. Wegrowe {\it The physical properties of the Hall current}, J. Appl. Phys. {\bf 129},  144501 (2021), https://doi.org/10.1063/5.0044912 
   
  \bibitem{SHE} M. Creff, E. Olive, and J.-E. Wegrowe, {\it Screening effect in Spin-Hall Devices}, Phys. Rev. B {\bf 105}, 174419 (2022).
  
  \bibitem{ArXiv_2024} D. Lacour, M. Hehn, Min Xu, J.-E. Wegrowe, {\it Injection of anomalous Hall current: the role of impedance matching}, arXiv: 2306.14226v2 (2023).
  
    \bibitem{DeGroot} See the sections {\it relaxation phenomena} and {\it internal degrees of freedom} in Chapter 10 of De Groot, S.R.; Mazur, P. {\it Non-equilibrium Thermodynamics}; North-Holland: Amsterdam, The Netherlands, 1962.
  
\bibitem{Rubi} D. Reguera, J. M. G. Vilar, and J. M. Rub\`i, {\it The mesoscopic Dynamics of Thermodynamic Systems}, J. Phys. Chem. B {\bf 109} (2005).
  
  \bibitem{Seitz} This equation can be found in the book edited by F. Seitz and D. Turnbull, {\it Galvanomagnetic and thermomagnetic effects in metals}, in the article of J.-P. Jan (equation (7.2) page 15), Solid State Physics, Academic Press Inc. Publishers, New York 1957. A more detailed analysis and a specific microscopic model can be found in T. McGuire and R. Potter,  {\it Anisotropic magnetoresistance in ferromagnetic 3D alloys}, IEEE Trans. Magn. 11, 1018 (1975).
  
 \bibitem{Matrix} J.-E. Wegrowe, D. Lacour, H.-J. Drouhin  {\it Anisotropic magnetothermal transport and spin Seebeck effect}, Phys. Rev. B {\bf 89}, 094409 (2014).
    
 \bibitem{Madon} B. Madon, M. Hehn, F. Montaigne, D. Lacour, and J.-E. Wegrowe.  {\it Corbino magnetoresistance in ferromagnetic layers : Two representative examples $Ni_{81}Fe_{19}$ and $Co_{83}Gd_{17}$ } Phys. Rev B (R) {\bf  98} 220405(R) (2018).
    
  \bibitem{Onsager_Diss} L. Onsager and S. Machlup, {\it Fluctuations and irreversible processes}, Phy. Rev. {\bf 91} 1505 (1953).
  
  \bibitem{Bruers} S. Bruers, Ch. Maes, K. Netocn\'y, {\it On the validity of entropy production principles for linear electrical circuits}, J. Stat. Phys. {\bf 129}, 725 (2007).
  
  \bibitem{MinDiss} L. Bertini, A. De Sole, D. Gabrielli, G. Jona-Lasinio, and C. Landim {\it Minimum Dissipation Principle in Stationary Non-Equilibrium Sates}, J. Stat. Phys. {\bf 116}, 831 (2004).    
        



  \end{thebibliography}
\end{document}